# Striving for Equity in Canadian Physics

Svetlana Barkanova[1, a)], Gwen Grinyer[2, b)], Juliette Mammei[3, c)],
Carolyn Sealfon[4, d)] and Anastasia Smolina[5, e)]

[1] *School of Science & Environment, Physics, Memorial University of Newfoundland, Grenfell Campus, 20 University Dr, Corner Brook, Newfoundland & Labrador, Canada*
[2] *Department of Physics, University of Regina, 3737 Wascana Parkway, Regina, Canada*
[3] *Department of Physics & Astronomy, University of Manitoba, 66 Chancellors Cir, Winnipeg, Manitoba, Canada*
[4] *Ronin Institute, 127 Haddon Pl, Montclair, New Jersey, United States of America*
[5] *Department of Medical Biophysics, University of Toronto, 101 College St, Toronto, Canada*

a) Corresponding author: sbarkanova@mun.ca
b) gwen.grinyer@uregina.ca
c) jmammei@physics.umanitoba.ca
d) carolyn.sealfon@ronininstitute.org
e) anastasia.smolina@mail.utoronto.ca

**Abstract.** We discuss a number of new initiatives and events since 2020 which we hope will contribute to advancement of equity issues within the physics community in Canada. A recent analysis of high-school data shows that men are still over-represented in high-school physics courses, and the fraction has not changed over a decade. Results from a national survey show that despite improvements over the years, the percentage of women and gender diverse physicists drops by around 35% between undergraduate students to those in a physics career. This decline is even more notable among Black, Indigenous, and people of colour (BIPOC) women and gender diverse physicists, whose representation drops by almost 60%. Several programs from the National Sciences and Engineering Research Council (NSERC) have been implemented in order to improve equity, diversity, and accessibility in STEM on a national level, most notably the Chairs for Women in Sciences and Engineering (CSWE) and Chairs for Inclusion in Sciences and Engineering (CISE) initiatives. It is crucial to maintain data collection and support existing as well as new EDI projects in future years as we work to build a more inclusive community of physicists in Canada.

The percentage of women studying physics in Canada goes down considerably between high school and full professor, from 40% female in high school to 14% as full professor [1, 2]. Even if all new hires were female (an unlikely scenario), it would take a decade or more to reach 'gender parity' in our country [3]. Some of us have had colleagues complain that hiring preferences for women or other equity-seeking groups is unfair to white and/or male students trying to enter the field. We point to the historic and ongoing systematic biases that have created the unbalance, and the many individual women who have been unfairly discriminated against in the past (and present). We respectfully remind our colleagues that we are trying to right a wrong on a societal scale, and propose that there are still significant barriers to entry and participation for women and other underrepresented groups.

In Ontario, one Province in Canada, most schools have 12 grades in elementary and high school, and then students can apply for post-secondary programs (see Fig. 1). Only about 1/3 of 12th grade students studying physics in Ontario are female [2]. This is a required course for post-secondary physics programs; the transition from Grade 11 to Grade 12 represents the single largest loss of potential women physicists. Action needs to be taken to improve the numbers of young women choosing physics courses after 10th grade (approximately 16 years of age), as this is a key timepoint that determines which future paths are available to them.

Work is needed to eliminate barriers for all equity-deserving groups, including women, gender-diverse people, disabled people, and BIPOC individuals in Canada. Distinguishing true allyship from tokenism must also be taken into account [4]. From the 2020 CanPhysCounts survey, we see that the percentage of women and gender diverse physicists drops by around 35% between undergraduate students to those in a physics career. This decline is even more notable among BIPOC women and gender diverse physicists, whose representation drops by almost 60% [1]. Note that this is the first data available on Black and Indigenous physicists, as well as those with non-binary gender identities; the vast majority of surveys have very limited identity categories. Among respondents with a disability, only 5% reported receiving full accommodations for their required needs at their place of work or study. BIPOC and non-heterosexual students were significantly more likely to identify as being disabled. In addition, we learned that the COVID-19 pandemic had disproportionately increased the caregiving responsibilities of women who are Black and/or Indigenous, and People of Colour more generally, exacerbating existing inequalities in a time of crisis [5]. Overall, students were found to be much more demographically diverse than working professionals, indicating how important it is to act early in order to retain the diverse physicists of tomorrow.

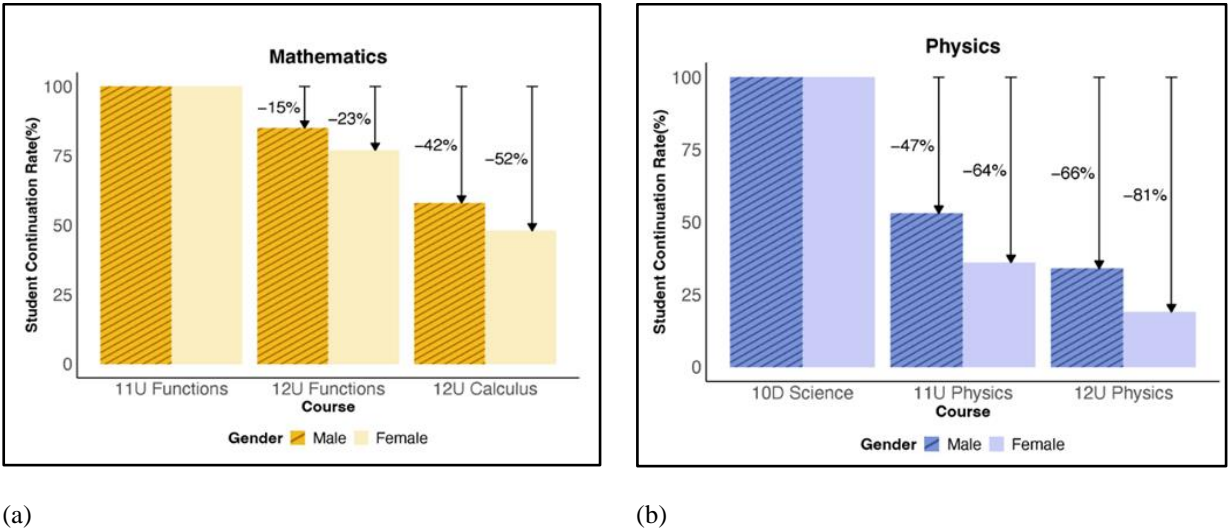

(a) (b)

**FIGURE 1.** *Student continuation rates in Math and Physics courses in Ontario (2007-2018). Leftmost columns show the number of students in required courses for Math (a) and Physics (b); the other columns show enrollment in later/elective courses.[2]*

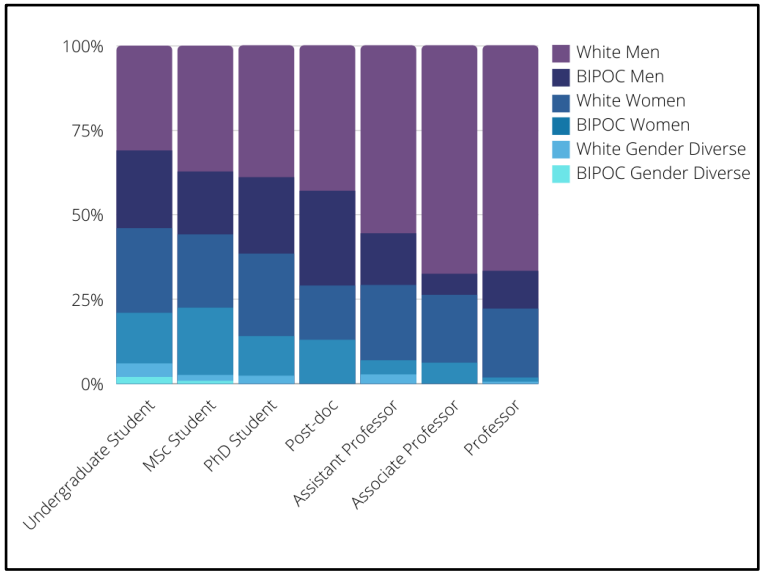

**FIGURE 2.** *Gender and racial identities across academic positions. Data from CanPhysCounts 2020 (N=2532); the BIPOC category includes Black Indigenous, and People of Colour, and the Gender Diverse category encompasses participants who identified as non-binary, Two-Spirit, genderqueer or gender-nonconforming, and those who self-described their gender [1].*

There are several Canadian groups working for equity in STEM at a National level, including the Canadian Association of Physicists Division for Gender Equity in Physics (CAP DGEP), the Canadian Conference for Undergraduate Women in Physics (CCUWiP), the CAP Equity Diversity and Inclusion Committee, professional organizations and large experimental collaborations which are adding codes of conduct and directly addressing EDI, the Canadian 2SLGBTQ+ in STEM Conference (Regina, 2024) and the Laurier Center for Women in Science. These groups work to support women and other equity-seeking groups to pursue the study of physics via opportunities to speak about their research, attend informative workshops and panel discussions, and network. Some experimental collaborations have incorporated EDI into their governance, including [nEXO](nEXO) and [EIC Canada](EIC Canada). In addition, most recent reports of the [Subatomic Physics](Subatomic Physics) and [Astro](Astro) Long Range Plans explicitly address EDI. The Arthur B. McDonald Canadian Astroparticle Physics Institute has implemented a Performance Measurement Plan ([PMP](PMP)) to assist the organization in measuring their progress towards greater inclusion across a national network. We need to take data regularly in order to track the efficacy of EDI initiatives and overall progress. This can be problematic for some groups, such as Indigenous and First Nations peoples in Canada, whose identification as Indigenous was used by the government to cause them harm. Great care needs to be taken to protect the anonymity of respondents, particularly for subgroups with the lowest numbers. We recommend following the guidelines of the National Science and Engineering Research Council (NSERC) and either not report any instances where there are fewer than 5 individuals in a group, or collapse them into a shared, larger group.

The primary federal funding body in Canada, NSERC, has several initiatives to increase equitable and inclusive access to funding opportunities and to foster a more equitable, diverse and inclusive post-secondary research ecosystem. For example, collaborations and individuals applying for NSERC research grants are now required to describe how they address EDI in their research groups [6], and are thus encouraged to acknowledge the barriers faced by underrepresented groups and examine how they work to help highly qualified personnel overcome these barriers. Unfortunately, the new program called "Dimensions" to facilitate EDI at the institutional level, as proposed in [6] by the Canadian National Science and Engineering Research Council, the Canadian Institutes of Health Research and the Social Sciences and Humanities Research Council (referred to as Tri-Agency), was canceled in 2023 due to funding constraints. However, NSERC can rely on the expertise of its seven regional Chairs for Women in Science and Engineering (CWSE) and Chairs for Inclusion in Science and Engineering (CISE) connected via national network [7] as it reviews and develops its EDI programs. In 2023, the highly-successful NSERC CWSE program [8] was expanded to include NSERC CISE, starting from the three inaugural Chairs in Atlantic Canada [9]. The Atlantic Chairs aim to reach underrepresented groups across the region, including women, Indigenous peoples, persons with disabilities, racialized groups, and members of the 2SLGBTQIA+ communities, with a particular focus on Indigenous and Francophone communities, African Nova Scotians and youth in rural and remote communities.

There is still a long way to go to achieve equity in physics. We are a large and diverse country with many intersecting identities, and aim to drive change from within our community in order to build a stronger foundation for generations of future physicists. We recognize that change happens slowly, but it is crucial to maintain and continue data collection and EDI initiatives aimed at improving physics culture across Canada. It is hard work, but we each have a role to play, and even making small changes in our departments and workplaces can lead to significant systemic changes over time.